\documentstyle[11pt]{article}
 
\def\pa{\partial}

\def\a{\alpha} 
\def\b{\beta} 
\def\d{\delta} 
\def\e{\epsilon}

\def\l{\lambda} 
\def\m{\mu} 
\def\n{\nu}

\def\s{\sigma}

\begin{document}

\begin{center}{\Large\bf Dimensionally Challenged Gravities}

S. Deser 

Physics Department, Brandeis University, Waltham, 
Massachusetts 02454, USA
\end{center}

\begin{quotation}
I review some ways in which spacetime dimensionality
enters explicitly in gravitation.  In particular, I recall
some unusual geometrical gravity
models that are constructible in dimensions different
from four, especially in D=3 where even ordinary Einstein 
theory is ``different", {\it e.g.}, fully Machian.
\end{quotation}

It is a pleasure to dedicate this little 
travelogue/catalogue of exotic gravity models to
John Stachel, whose loyalty to the D=4 Einstein
cause is too steadfast to be subverted by reading it.

Once unleashed by general relativity, dynamical 
geometry has become a fertile playground for 
generalization in many directions beyond Einstein's
D=4 Ricci-flat choice.  This trend has intensified 
with string theory, where D=10 is normal as are 
(higher curvature power)
corrections to the
Einstein action.  There are many other reasons to 
study different dimensions; here is one: As I became
aware, thanks to John, Einstein already foresaw
\cite{001} the potential danger of letting geometry
be at the mercy of field equations, in particular
worrying about spaces with closed timelike curves,
but also optimistically hoping that they would be
forbidden in ``physically acceptable" matter contexts
(this is {\it not} a tautology since acceptable means
having decent stress tensor).  Although the 
best-known examples, such as G\"{o}del's universes
\cite{002}, fall in this class, it is in the 
simpler setting of D=3 (``planar") gravity that they
have recently been studied on an industrial scale
\cite{003}, 
and have yielded Einstein's hoped-for taboo in
a clear way.  More generally, one can learn about D=4
Einstein's virtues from studying different D's,
and the different sorts of models
they support.  What is more, we are very likely to be
embedded in a world, which, if it has any classical
geometry at all, is likely to have as many as eleven
dimensions!  In this short excursion, I can only
point out some recent examples of theories that I
have been involved with directly; equations and further
references will have to be found in the citations.

Let us begin with some remarks about ordinary Einstein 
theory in the smaller worlds of D$<$4.  One does not 
normally think of the
curvature components as being dimension-dependent, 
but we all know that in D=3, Einstein and Riemann tensors 
have the same number of components and indeed
are equivalent,  since $G^\m_\n = \frac{1}{4} \e^{\m\a\b}
\e_{\n\l\s} R_{\a\b}~^{\l\s}$. Strangely,
it was a long time before the import of this was
appreciated: that outside sources, spacetime is flat!
More precisely it is locally, but not globally, flat.
Philosophically, D=3 Einstein theory
presents the Machian dream in its 
purest form: there are no gravitational excitations, so
geometry is entirely -- and locally -- determined by
matter.  There is a field-current identity: Riemann
(being Einstein) equals stress tensor.  So the picture
that emerges is that this planar world 
consists of patches of
Minkowski space glued together at the sources (most
simply discrete point particles, representing
parallel strings in a D=4 Einstein world).  The 
1-particle conical space
solution is amusing enough \cite{004} but things 
really get to be fun for two or more stationary or,
better still, moving ones \cite{005}.  If a
cosmological constant is present, it's even more
fun as the patches are constant curvature spacetimes
\cite{006}. In that case (for negative cosmological 
constant) it is also possible to have
black holes by suitable identifications of points, 
\cite{007}.
Time-helical structures, requiring identification of
times in a periodic way (as well as the space
gluings)arise for stationary, rotating, solutions and
lead to the whole gamut of possible closed
timelike curves and, as mentioned,
a clear arena to examine whether they can be physically
generated.  But D=3 can be more amusing still, for
it permits (as does any odd-dimensional space) the
construction of different invariants, the 
Chern--Simons (CS) terms.  These are the gravitational
analogs of the simple electrodynamics (or Yang--Mills)
$\int A \wedge F$ structures that in turn arise from
the next higher dimensional topological invariants
such as $F_{\m\n}~^\star F^{\m\n} \equiv
\pa_\m (\e^{\m\n\a\b} A_\n F_{\a\b})$ in the 
abelian context. Here we 
have the Pontryagin invariant $R^\star R$ 
instead.  Varying
these gravitational CS terms with respect to the
metric leads to a tensor, because the integral 
(if not the CS integrand) is gauge invariant.  
In D=3, this is the famous Cotton tensor, $C^{\m\n} 
\equiv \e^{\m\a\b} D_\a (R^\n_\b - \frac{1}{4} 
\, \d^\n_\b R)$, discovered long before general 
relativity \cite{008}; $C^{\m\n}$ is the conformal
tensor in D=3, replacing the (identically vanishing)
Weyl curvature.  It is a symmetric, traceless,
identically conserved quantity, although it 
superficially seems to be none of these.  
Its interest lies not so much for generating
a theory of gravity in its own right (it could
at best only couple to traceless sources) but as an
added term to the Einstein one.  Being of third
derivative order, it has a coefficient with relative
inverse length or
mass dimension (in Planck units) to that of 
the Einstein action.  This mass is in fact that
of small excitations (of helicity $\pm$2) of the
metric about flat space: adding CS has restored a
degree of freedom absent in either $R$ or CS alone.
This combined theory \cite{009}, called topologically
massive gravity (TMG) for obvious reasons, has many 
other wondrous properties and unsolved aspects. 
First, despite being a higher derivative theory,
it has no unitarity or ghost problems; it may even
be finite as a quantum theory, although that is 
still an open mathematical problem \cite{010}.  If
so, it might really have some lessons for us, for
it would be unique in this respect amongst truly
dynamical gravity models without ghosts (unlike 
four-derivative theories) but with a 
dimensional coupling constant; pure Einstein D=3
theory is renormalizable but that doesn't count --
it is non-dynamical \cite{011}.  Second, TMG, at least
in its linearized guise \cite{012} acts to turn its
sources into anyons; that is, a particle can acquire
any desired spin simply by coupling to TMG.  But,
thirdly, no-one has succeeded as yet in finding the 
simplest possible, ``Schwarzschild" solution to the
nonlinear model, {\it i.e.}, a circularly symmetric
time-independent (we don't even know if there's
a Birkhoff theorem) exterior geometry that obeys the
$G_{\m\n} + m^{-1} C_{\m\n} = 0$ equations.  
Although CS-like terms
can be constructed for higher odd-D spaces, they 
have not been studied much because they have no
linearized, kinematical,
effects beyond D=3 because they are of higher powers
in an expansion about flat space.  There are both
strong similarities and differences between TMG 
and its spin 1 counterpart, topologically massive 
Yang--Mills theory. The most striking
difference is that in the quantum theory, the
coefficient of the CS term in ``TM--YM" must
necessarily be quantized \cite{009}, but 
not that of TMG \cite{013}.

But the twists in D=3 gravity do not stop there:
there is yet another ``CS-ness" present.  Once
it is noted that, in Einstein gravity, spacetime is
flat outside sources, one realizes that
this is just the same
as what happens to abelian or nonabelian vector
fields in their pure CS models: the field equations
are just $^\star F^\m \equiv \frac{1}{2}\: \e^{\m\n\a}
F_{\n\a} = 0$, so that the field ``curvature" also
vanishes here 
[in both cases, the full ``curvatures" are proportional
to currents, $^*\! F^\m = J^\m$].  Indeed, there is an
equivalence (except for some interesting find print)
between the two models and one can formally 
recast the Einstein action and equations into
non-abelian vector field CS form in terms of the
dreibein and spin connections.  So this is yet
another vast subject straddling two ostensibly
different types of theory; for a review see
{\it e.g.} \cite{014}.

We will not descend much to D=2, another
well-studied subject \cite{015}, because there is
no Einstein gravity there at all: only the Ricci
scalar is non-vanishing, being the ``double-dual"
of Riemann, while the Einstein tensor vanishes
identically.  As usual, D=2 is different from all
other dimensions in this respect (it is also here
that Maxwell theory ceases to have excitations);
some sort of additional scalar field is required
to assure the Hilbert action from just being a 
dull Euler topological invariant and this departs
from the realm of pure geometry.

What about dimensions beyond D=4?  This becomes a
generic area where the differences from D=4 are
more quantitative than qualitative.  
Still, there are some amusing
points to be noted.  For example, consider the 
Gauss--Bonnet invariant $R^\star R^\star$,
defined in D=4.  There, it is a total divergence
and hence irrelevant to field equations.  However,
in higher dimensions, it can still be defined by
writing it out in terms of metrics; for example we
all know it is proportional to the combination
$(R^2_{\m\n\a\b} - 4 R^2_{\m\n} + R^2 )$.  As
a Lagrangian, it is seemingly dangerous to unitarity
of excitations because of its fourth derivative
order.  In fact, there is no danger, 
because (say about flat space) this
combination is a total divergence in its leading
quadratic order in $h_{\m\n} \equiv (g_{\m\n}
- \eta_{\m\n})$ in {\it any} D.  Thus, this is a
``safe" class of alternative actions, say when 
added to Einstein's. Some of their solutions,
{\it e.g.}, Schwarzschild-like ones, have been
studied to see whether they are better or still
unique.  For example, there can be cosmological
solutions without an explicit cosmological
constant \cite{016}, which is not necessarily
a good thing, physically.

In quantum field theory, a powerful tool has been
the ``large $N$ limit" of Yang--Mills theory 
in which the number of flavors (internal
degrees of freedom: $A^a_\m$, $a = 1 \ldots N$) is
sent to infinity.  The equivalent in gravity, 
rather naturally, the dimensionality D of
spacetime, over which the ``internal" index $a$
of the vielbein $e^a_\m$ ranges.  As far as I know,
the literature here consists of but one brave
paper \cite{017}, which however did not get far.
This seems to me a worthy subject of study also by
classical relativists, who have instead mostly 
considered what is in some ways the opposite,
ultra-local,  limit 
\cite{018}.  There must be some 
simplifying aspects as the number of degrees of
freedom $\frac{1}{2} D(D-3)$, rises quadratically, 
and the Newtonian potential that enters in the 
Schwarzschild metric behaves as $r^{-(D-3)}$.

There are also auxiliary quantities that are
interestingly dimension-dependent; we encountered
some of them in current work on D=11 supergravity
\cite{019}.
I will not transgress further into the superworld
here, except to say that it is absolutely amazing 
that ~a) Einstein gravity always has a ``Dirac
square root" for all D$\leq$11, {\it i.e.}, can
always be consistently supersymmetrized without 
the need for higher spins or more than one graviton, 
and ~b) that this possibility
stops \cite{020} at D=11 and
~c) that cosmological terms are allowed for all
D$<$11, but forbidden \cite{021} at D=11.  
I certainly do not know of any
``Riemannian" differences between D=11 and D=12 for 
example!
In our case, we needed to find a basis of monomial
local invariants made up from Riemann or Weyl 
tensors at a given quartic order. Their number 
had a natural ``generic" bound \cite{022} at
D=8, whereas D=4 is
always the most degenerate dimension (I drop
D$<$4 here).  This sort of property is also of
importance when studying gravitational trace
anomalies \cite{023} in which 
identities arising from
antisymmetrizing an expression over more indices
than there are dimensions essentially generate
all the strange looking identities between
tensors, such as $C^{\m\l\a\b}C^\n_{\l\a\b}
= \frac{1}{4} \, g^{\m\n} C^2$ in D=4,
from antisymmetrizing expressions
such as $C^{\m\n}_{[\m\n} C^{\a\b}_{\a\b}
X^\l_{\l ]}$, where $X$ is arbitrary and brackets
indicate antisymmetizations over 5 indices.  
In the anomaly context we are
actually interested, using so-called dimensional
regularization, in spaces of dimension differing
from an integer by an infinitesimal parameter, still
another unlikely departure from D=4, but one that 
has its own unlikely set of geometrical rules.  
In conclusion, I have
tried to indicate that the list of
interesting, useful and even important consequences
to be drawn from excursions away from our favorite
Einstein action in its D=4 world is 
substantial and by no means complete.

This work was supported by NSF grant PHY 93-15811. 
I am grateful to my collaborators in our explorations
of the areas discussed here.

\end{document}